\documentclass[aps,prl,twocolumn,showpacs,amssymb,amsmath]{revtex4}
\usepackage{graphicx}
\usepackage{epsfig}
\usepackage{bm}

\begin{document}

\title{Fractionalization of a flux quantum in a one-dimensional parallel Josephson junction 
array with alternating $\pi$ junctions}
\author{Mahesh Chandran$^1$}
\email{mchandran@tifr.res.in}
\author{R. V. Kulkarni$^2$}
\email{rahul@nec-labs.com}
\affiliation{$^1$Tata Institute of Fundamental Research, Homi Bhabha Road, Mumbai, 400 005, 
INDIA. \\
$^2$Department of Physics, University of California, Davis, CA 95616.}
\altaffiliation{Present Address:  NEC Labs, 4 Independence Way, Princeton, NJ 08540.}

\date{\today}

\begin{abstract} 

We study numerically and analytically the properties of a one-dimensional array of parallel
Josephson junctions in which every {\em alternate} junction is a $\pi$ junction. In the
ground state of the array, each cell contains spontaneous magnetic flux
$\Phi\!\leq\!\Phi_{0}/2$ which shows {\em antiferromagnetic} ordering along the array. We
find that an externally introduced $2\pi$-fluxon $\Phi_{0}$ in such an array is unstable
and fractionalizes into two $\pi$ fluxons of magnitude $\frac{1}{2}\Phi_{0}$. We attribute
this fractionalization to the degeneracy of the ground state of the array. The magnitude of
the flux in the fractional fluxons can be controlled by changing the critical current of
the $\pi$ junctions relative to the 0 junctions. In the presence of an external current,
the fluxon lattice in the antiferromagnetic ground state can be depinned. We also observe a
novel resonant structure in the $V$-$I$ characteristics above the depinning current due to
the interaction between the fluxon lattice and the array.

\end{abstract}

\maketitle

One of the exciting developments in the field of Josephson devices is the fabrication
of the three terminal controllable Josephson junction~\cite{morpurgo}. The supercurrent
through a Josephson junction is given by $I=I_{c}\sin(\Delta\phi)$ where $\Delta\phi$
is the gauge-invariant phase difference between the superconductors, and the critical
current $I_{c}$ depends upon the junction geometry, normal state resistance $R_{n}$ and
the temperature $T$. Morpurgo {\it et al.}\cite{morpurgo} showed that the supercurrent
through a superconductor-metal-superconductor (SNS) junction changes on passing a
control current through the normal metal. For such a junction, the supercurrent
$I_{}\propto\sin(\Delta\phi+\chi)$ where the additional phase difference $\chi$ is
dependent on the current through the normal metal. Further theoretical work showed that
in the diffusive limit of the junction, the additional phase factor $\chi$ can be made
$\pi$, thus reversing the direction of the supercurrent with respect to the phase
difference $\Delta\phi$~\cite{volkov, wilhelm}. Josephson junction with $\chi\!=\!\pi$ 
is referred as the $\pi$ junction (we use the term 0 junction for the Josephson 
junction for which $\chi\!=\!0$). The $\pi$ junction has now been realized in 
several experiments~\cite{baselmans, shaikh, comment1}. The fabrication of such tunable
junctions has opened immense possibilities for new applications, as demonstrated
recently by the development of controllable $\pi$-SQUID~\cite{baselmans1}.

The next natural step in this field would be to consider Josephson junction array (JJA)
containing $\pi$ junctions. Theoretically, Kusmartsev~\cite{kusum} considered a loop
containing an odd number of $\pi$ junctions and showed that the loop contains spontaneous
magnetic flux in the ground state. In the continuum limit, the long Josephson junction with
alternating critical current density have been studied which shows self-generated magnetic
flux~\cite{mints,gold}. Recent studies of JJAs with $\pi$ junctions~\cite{mahesh,kulkarni}
have shown some novel features arising out of the interplay between 0 and $\pi$ junctions.
Moreover, JJA is a unique system which provides experimental realizations of several
interesting physical phenomena, some examples of which are field induced superconductor to
insulator transition~\cite{zant}, Aharonov-Casher effect~\cite{elion}, coherent emission of
radiation~\cite{barb}. One is then led to ask as what new physical phenomenon exist in the
1D JJA containing $\pi$ junctions.

In this paper, we study numerically and analytically a new class of 1D JJA: an array
of parallel Josephson junctions in which every {\em alternate} junction is a
$\pi$ junction. The ground state contains spontaneous magnetic flux in each cell and
are ordered {\em antiferromagnetically} along the array. We find that a quantum of
flux (fluxon) with a $2\pi$ kink in the phase is unstable in such an array, and
fractionalizes into two spatially separated $\pi$ kink fluxons. We also calculate
the $V$-$I$ characteristics which shows a novel structure due to resonant
interaction between the moving antiferromagnetic fluxon lattice and the linear waves
emitted by the array.

Consider a 1D array of parallel Josephson junctions containing alternate $\pi$- and
0 junctions (see inset Fig. 1). The Hamiltonian for this system is
\begin{equation}
\frac{H}{E_{J}}=\sum_{i=0}^{N-1}\frac{1}{2}\left(\frac{\partial\phi_{i}}{\partial
t}\right)^{2} + \frac{\lambda_{J}^{2}}{2}(\phi_{i+1}-\phi_{i})^{2} +
[1-(-1)^{i}\cos\phi_{i}],
\end{equation} 
where $\phi_{i}$ is the gauge-invariant phase difference across the $i$-th junction. The
periodic boundary condition is imposed at the two ends of the array such that
$\phi_{0}\!=\!\phi_{N}$ ($N$ is assumed to be even). In Eq.(1), the first term represents the
charging energy and the second term is the energy of the induced magnetic field due to finite
self-inductance of the cell (the effect of mutual inductance between the cells is neglected).
The last term represents the energy associated with the Josephson currents. The prefactor for
the $\cos\phi_{i}$ term alternates in sign for odd ($\pi$-) and even (0-) junctions. The
Josephson coupling energy $E_{J}\!=\!I_{c}\Phi_{0}/2\pi$ where $I_{c}$ is the critical current
of a single junction. The time $t$ is in the units of inverse plasma frequency
$\omega_{P}^{-1}\!=\!\sqrt{\frac{\Phi_{0}C}{2\pi I_{c}}}$, where $C$ is the averaged
capacitance per unit area of the junction. The effective Josephson penetration depth is given
by $\lambda_{J}\!=\!(\frac{\Phi_{0}}{2\pi L_{0}I_{c}})^{1/2}$ where $L_0$ is the self
inductance of a single cell. The $\lambda_J$ determines the screening strength of the array,
and is related to the SQUID parameter $\beta_{L}\!=\!\lambda_{J}^{-2}$.

From Eq.(1), the equation of motion for $\phi_i$ is 
\begin{equation}
\frac{d^{2}\phi_{i}}{dt^{2}} + \alpha \frac{d\phi_{i}}{dt} + (-1)^{i} \sin\phi_{i} +
\gamma = \lambda_{J}^{2}(\phi_{i+1}+\phi_{i-1}-2\phi_{i}), 
\end{equation} 
where a dissipative term $\alpha\frac{d\phi_{i}}{dt}$ is also added~\cite{barone}. The
coefficient $\alpha\!=\!\beta_{c}^{-1/2}$, where $\beta_{c}\!=\!2\pi
R_{n}^{2}I_{c}C/\Phi_{0}$ is the McCumber parameter. The parameter
$\gamma\!=\!I_{ext}/I_{c}$ represents the external current through the junction.  For the
numerical simulation, Eq.(2) is integrated using the fourth order predictor-corrector method.  
The consistency of the steady state solutions was checked using different initial
configurations of $\phi_i$'s. The magnetic flux in the $i^{th}$ cell is defined as
$2\pi\frac{\Phi_{i}}{\Phi_{0}}=-(\phi_{i+1}-\phi_{i})$. We remark that for the case where
all junctions are 0 junctions (henceforth referred to as the 0-JJA), Eq.(2) is the
discrete perturbed sine-Gordon equation, and have been studied
extensively~\cite{ustinov1,ustinov2}. First, we consider the results from the numerical
simulation.

\begin{figure}[hbt] 
\includegraphics[width=210pt]{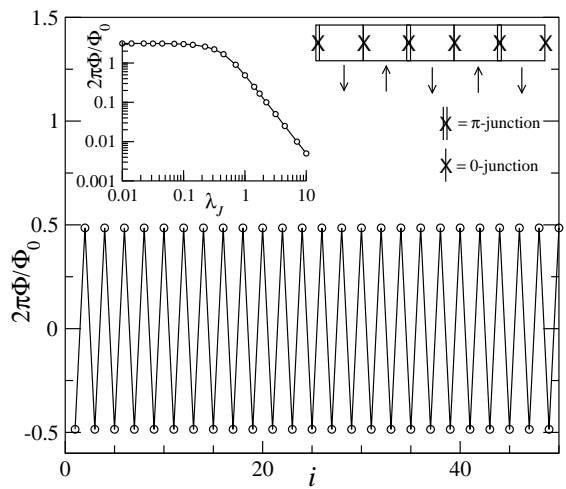}
\caption{The self-induced magnetic flux $2\pi\Phi_{i}/\Phi_{0}$ along the array in the
ground state. The $\lambda_{J}\!=\!1.0$ and $N\!=\!100$ (only half the array is shown for
clarity). The right inset shows the array geometry and the arrows represents the
antiferromagnetic ordering of the magnetic flux indunced in the cell. The left inset shows
the dependence of $2\pi\frac{|\Phi_{i}|}{\Phi_{0}}$ on $\lambda_{J}$.} 
\end{figure}

Figure 1 shows the ground state flux configuration $2\pi\Phi_{i}/\Phi_{0}$ for $N\!=\!100$ and
$\lambda_{J}\!=\!1.0$. The self-induced magnetic flux $\Phi_i$ changes sign across neighboring
cells with $|\Phi_{i}|=\Phi$ remaining constant. Such a configuration of $\Phi_i$ is
reminiscent of the ground state in 1D classical Ising model with {\em antiferromagnetic} (AF)
coupling, as depicted schematically in the inset of Fig. 1. Therefore, we call this array the
antiferromagnetic JJA (AFJJA). The AF ordering of $\Phi_i$ (and hence, $\phi_i$) implies that
the self-induced screening currents in neighboring cells are oppositely oriented. The
magnitude of the flux $\Phi$ in a cell depends on the screening strength $\lambda_J$ as shown
in the Fig. 1(inset). With increasing $\lambda_J$, the magnetic flux in the neighboring cells
tends to overlap, and $\Phi\rightarrow 0$ as $\lambda_{J}\rightarrow\infty$. In the strong
screening limit, $\lambda_{J}\rightarrow 0$ and $\Phi\rightarrow\frac{\Phi_{0}}{2}$.

Next, we consider the consequence of introducing an external fluxon in
AFJJA\cite{comment3}. In the 0-JJA, a fluxon corresponds to a $2\pi$-kink in the phase
profile $\phi(x)$ and the magnetic field ($\propto \frac{\Delta\phi}{\Delta x}$) is
spatially localized on the length scale $\lambda_J$.  Figure 2(a) shows the steady state
profiles of $2\pi\Phi_{i}/\Phi_{0}$ and $\phi_i$ in AFJJA in the presence of a
$2\pi$-fluxon. The self-induced magnetic field of the AF ground state has been subtracted
from $2\pi\Phi_{i}/\Phi_{0}$. We find that a $2\pi$-fluxon is unstable in the AFJJA and
{\em fractionalizes into two spatially separated fluxons}, each carrying half the quantum
of flux. Also, each of {\em the fractional fluxons is a $\pi$-kink} in $\phi_{i}$. The
magnetic field around the fractional fluxon decays as $\exp(-x/\lambda_{eff})$ where
$\lambda_{eff}\approx\! 2\lambda_{J}$. This should be compared with the $2\pi$-fluxon in
the 0-JJA [Fig. 2(b)] where $\lambda_{eff}\approx\lambda_{J}$. The increase in
$\lambda_{eff}$ in AFJJA is a consequence of the magnetic flux in the AF ground state.

\begin{figure}[hbt]
\includegraphics[width=210pt]{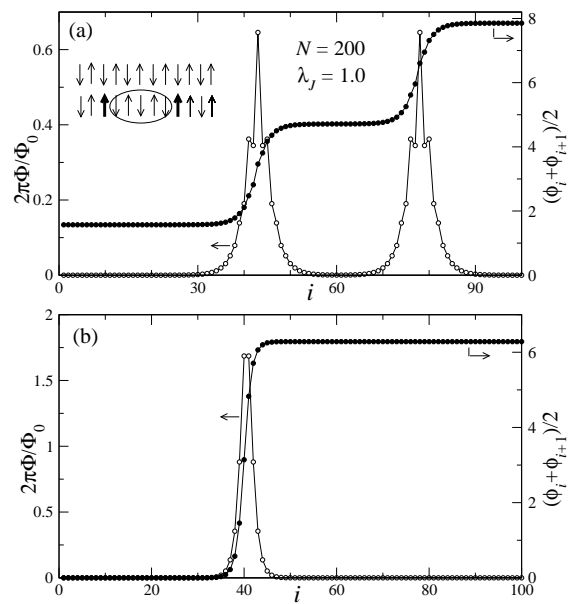} 
\caption{(a) The magnetic flux $2\pi\frac{\Phi_{i}}{\Phi_{0}}$ and the average phase
profile along the array in (a) AFJJA and (b) 0-JJA, in the presence of a $2\pi$-fluxon.
The background AF ground state has been subtracted from $2\pi\frac{\Phi_{i}}{\Phi_{0}}$ in
(a). The schematic in inset (a) shows how the fractional fluxons (thick arrows)
interpolate between the two degenerate ground state of $\Phi_i$'s.} 
\end{figure}

It is possible to vary the magnitude of the magnetic flux in each fraction by changing
$i_{c}^{*}\!=\!I_{c}^{\pi}/I_{c}^{0}$ where $I_{c}^{\pi}$ and $I_{c}^{0}$ are the critical
currents of the $\pi$ and 0 junctions, respectively\cite{comment2}. Figure 3(a) shows the
spatial profile of the fractional fluxons for $i_{c}^{*}=0.8$ and $\lambda_{J}=1.0$. The
phase change across the fractional fluxons is {\em not} $\pi$ but is dependent on the value
of $i_{c}^{*}$. The total phase change across both the fractions is always $2\pi$, as
required by the flux conservation. Figure 3(b) shows the magnitude of the integrated flux
$2\pi\frac{\Phi_{T}}{\Phi_{0}}$ in each fractional fluxons as a function of $i_{c}^{*}$ for
$\lambda_{J}=1.0$. The fractionalization occurs for
$i_{cl}^{*}\!<\!i_{c}^{*}\!<\!i_{cu}^{*}$, where $i_{cl}^{*}$ and $i_{cu}^{*}$ are the two
critical values. The slope $\frac{d\Phi_{T}}{di_{c}^{*}}$ at the critical values
$i_{cl}^{*}$ and $i_{cu}^{*}$ appears to diverge, suggesting a transition between the
fractionalized state and the single fluxon state. In experiments, the magnitude of the flux
at the center of the fluxon $\Phi_m$ can be measured more easily. Figure 3(b) shows the
behavior of $\Phi_{m}(i_{c}^{*})$.

In Fig. 4, we show the numerically obtained parameter space $\lambda_{J}$-$i_{c}^{*}$. We have
assumed that $i_{c}^{*}$ can be varied independent of $\lambda_J$. The region of fractional
fluxons is bounded by $i_{cl}^{*}(\lambda_{J})$ and $i_{cu}^{*}(\lambda_{J})$. It is easy to
understand the absence of fractional fluxons in the limit $i_{c}^{*}\!\rightarrow\!0$ since
the array becomes a 0-JJA (with lattice constant twice the original array) which allows only
$2\pi$-fluxons. In the opposite limit $i_{c}^{*}\!\rightarrow\!\infty$ such that
$I_{c}^{0}\rightarrow 0$ and $I_{c}^{\pi}$ is finite, there are two $\pi$ junctions in each
cell and the array can be shown to be equivalent to the 0-JJA, and the fractionalization is
again not expected. In obtaining the parameter space in Fig. 4, $I_{c}^{0}$ is assumed to be
finite and fixed which leads to fractionalization for $\lambda_{J}\!<\!0.7$ even as
$i_{c}^{*}\rightarrow\infty$.

\begin{figure}[hbt]
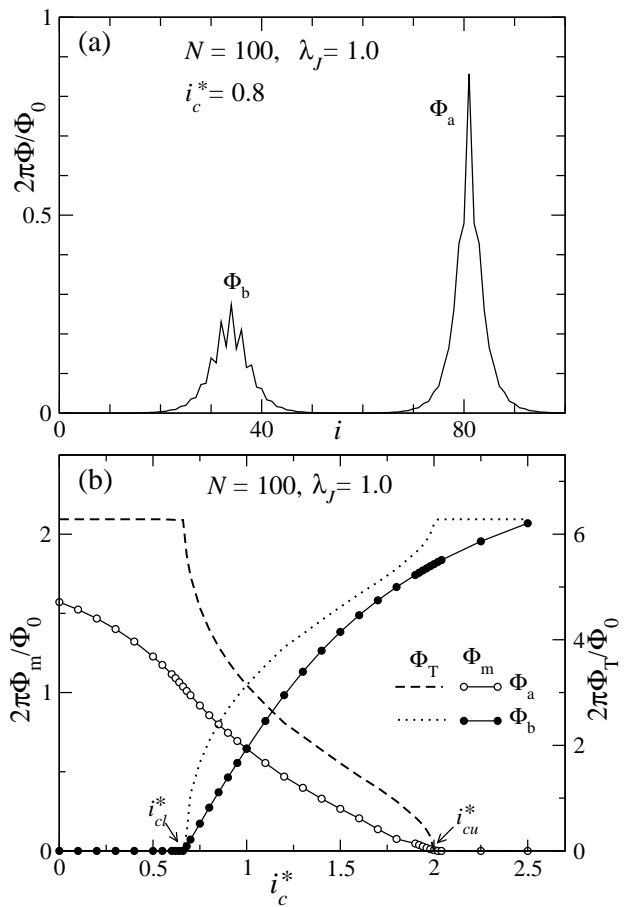

\includegraphics[width=210pt]{fig3a} 
\includegraphics[width=210pt]{fig3b} 
\caption{(a) The spatial profile of the fractionalized external fluxon for
$i_{c}^{*}\!=\!0.8$ and $\lambda_{J}\!=\!1.0$. The magnitude of flux in the two fractions are 
not equal. (b) The integrated total flux $2\pi\frac{\Phi_{T}}{\Phi_{0}}$ in the two
fractional fluxons (represented by dashed and dotted lines) as a function of
$i_{c}^{\ast}=I_{c}^{\pi}/I_{c}^{0}$. Also shown is the maximum flux
$2\pi\frac{\Phi_{m}}{\Phi_{0}}$ at the center of the two fractional fluxons $\Phi_a$ and
$\Phi_b$ (represented by symbols).}
\end{figure}

The simulation results discussed above can also be understood analytically. Consider the
case of $i_{c}^{*}\!=\!1$. Define
\begin{equation} 
\phi_{2m+1} = u_{m}, \hspace{0.2in} {\rm and} \hspace{0.2in} \phi_{2m} = v_{m}, 
\end{equation} 
where $m = 0, \frac{N}{2}-1$. Thus, $u_m$ and $v_m$ are the gauge
invariant phase differences across the $\pi$ and 0 junctions,
respectively. In the absence of any external fluxon, the $u_m$ and $v_m$ are invariant on 
translation by the lattice vector along the array. Hence, substituting $u_{m}\!=\!u$ and 
$v_{m}=v$, Eq.(2) becomes 
\begin{equation} 
\sin u = \sin v = 2\lambda_{J}^{2} (u - v). 
\end{equation} 
There are two trivial solutions of the Eq.(4): $u\!=\!v\!=\!0$ and $u\!=\!v\!=\!\pi$. The 
non-trivial solution of Eq.(4) is given by 
\begin{equation}
u = \pi - v, \hspace{0.2in} {\rm and} \hspace{0.2in} \sin v = 2\lambda_{J}^{2}(\pi - 2v). 
\end{equation}
For a given value of $\lambda_{J}$, the quantities $v$ and $u$ can be calculated
graphically from Eq.(5). It can be easily verified that the non-trivial solution is the
ground state for any finite $\lambda_{J}$. The magnetic flux in the cell is given by $\Phi
= \pm (u-v)\frac{\Phi_{0}}{2\pi}$, where the $+$ and $-$ sign is for the cell to the left
and the right of the $\pi$ junction, respectively. Thus, the magnetic flux alternates in
sign along the array. The values of $u$, $v$ and $|\Phi|$ obtained from Eq.(5) are in
excellent agreement with the numerically obtained values.  In the strong screening limit
$\lambda_{J}\rightarrow 0$, $u\!=\!\pi$ and $v\!=\!0$, and the flux in each cell attains
the maximum value $\Phi_{0}/2$. In the limit $\lambda_{J}\rightarrow\infty$, the solution
of Eq.(5) is $u\!=\!v\!=\pi/2$ which is degenerate to the trivial solutions of Eq.(4), and
the ground state contains no spontaneous magnetic flux.

\begin{figure}[hbt]
\includegraphics[width=210pt]{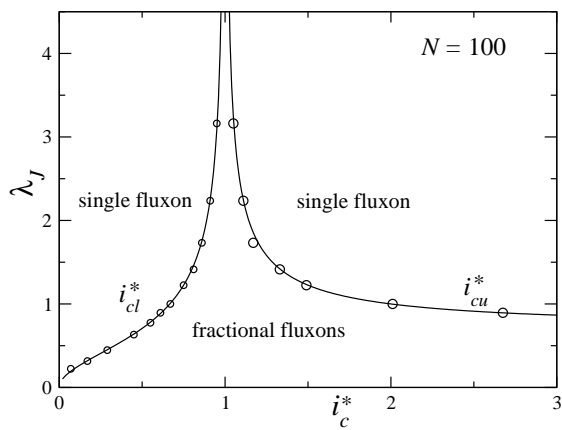} 
\caption{The parameter space $\lambda_{J}$-$i_{c}^{\ast}$. The circles are the numerically
obtained values whereas the full lines are the analytical result.}
\end{figure}

To understand the fractionalization of a $2\pi$-fluxon, we note that the AF ground state
solution of Eq.(4) is two-fold degenerate. If $\phi\!=\!\{u,v\}$ obtained from Eq.(5) is
one solution, the other solution is obtained by translating $\phi$ by one lattice
constant. Thus, the other solution is $\phi'\!=\!\{u',v'\}$, where $u'\!=\!v\!+\!\pi$ and
$v'\!=\!u\!+\!\pi$. The degeneracy of the ground state has an important implication: the
elementary excitation for the array is a kink (or domain wall) in $\phi_i$ which
interpolates between the two degenerate ground states $\phi$ and $\phi'$. It can be easily
verified that the phase change across the `kink' is $\pi$ and corresponds to an additional
flux $\Phi_{0}/2$ in the array. In the presence of a $2\pi$ fluxon, the energy
minimization leads to two $\pi$-kinks which are separated by the degenerate ground
states. This is shown schematically in the inset of Fig. 2(a). Fractionalization of a
$2\pi$ fluxon in AFJJA is thus a consequence of the degeneracy of the ground state. A
similar phenomenon is observed in polyacetylene~\cite{su} and certain field
theories~\cite{jackiw,rajaraman} where fractional topological excitations occur and are
related to the ground state degeneracy.

Equations (4) and (5) can be extended for the case $i_{c}^{*}\neq 1$. We find that the AF
ground state is stable and fluxon fractionalization occurs for $i_{c}^{*}>1$ when
$\lambda_{J}\!<\!\lambda_{Ju}\!=\!\frac{1}{\sqrt{2}}(1-\frac{1}{i_{c}^{*}})^{-1/2}$, and
for $i_{c}^{*}\!<\!1$ when
$\lambda_{J}\!<\!\lambda_{Jl}\!=\!\frac{1}{\sqrt{2}}(\frac{i_{c}^{*}}{1-{i_{c}^{*}}})^{1/2}$.
This is plotted in Fig. 4 and is in good agreement with the numerically obtained
behavior of $i_{cu}^{*}(\lambda_{J})$ and $i_{cl}^{*}(\lambda_{J})$. Further details of
the analytical calculations will be given elsewhere.

Next, we study the dynamical properties of the AFJJA. We restrict the analysis to the case
$i_{c}^{\ast}\!=\!1.0$ and in the absence of any external fluxon. The $V$-$I$ curve is
obtained by sweeping $\gamma$ in small steps, and calculating
$V\!=\!\alpha\langle\frac{d\phi}{dt}\rangle$ in the steady state ($V$ is in units of
$R_{n}I_c$). Recall that for the 0-JJA, all junctions switch from the superconducting state
to the normal state simultaneously at $\gamma\!=\!1$. In the AFJJA, the magnetic flux in the 
ground state alters this behavior significantly.

Figure 5 shows the $V$-$I$ curve for $N\!=\!30$ and $\lambda_{J}\!=\!1$. The $\alpha\!=\!0.1$
for the rest of the discussion below\cite{comment4}. The important feature of the $V$-$I$
curve is the appearance of a plateau in $V$ above a depinning current $\gamma_c$. The
transition to the normal state occurs at a higher current $\gamma_n$. By analyzing the
spatio-temporal dynamics of fluxons on the voltage plateau, we find that the
interpenetrating lattice of the flux ($\Phi$) and the antiflux ($-\Phi$) move in the 
opposite directions which appears as a stationary wave of breathing flux-antiflux
pairs~\cite{likharev}. This is evident from the 2D space-time plot of
$2\pi\frac{\Phi_{i}}{\Phi_{0}}$ which is shown for a section of the array in the inset of
Fig. 5. The $\frac{\Phi}{\Phi_{0}}$ in each cell oscillates between the positive and the
negative values, and is in antiphase with the neighboring cells. Thus, at any instant of
time, the total flux in the array is zero as expected from the ground state. We also find
that for $\lambda_{J}\!>\!1$, a linear flux flow regime appears before the plateau, whereas
for small $\lambda_J$, the $V$ shows a sharp step as shown in the inset of Fig. 5. The
$\lambda_{J}$ dependence of the width of the voltage plateau
$(\Delta\gamma)_{s}=\gamma_{n}-\gamma_{c}$ is shown in Fig. 6(a). The $(\Delta\gamma)_{s}$ is
non-monotonic and is maximum for $\lambda_{J}\approx 2$. Below a cut-off
$\lambda_{J}^{*}\approx 0.45$, the plateau in $V$ disappears and $\gamma_{c}=\gamma_{n}$.

\begin{figure}[hbt]
\includegraphics[width=210pt]{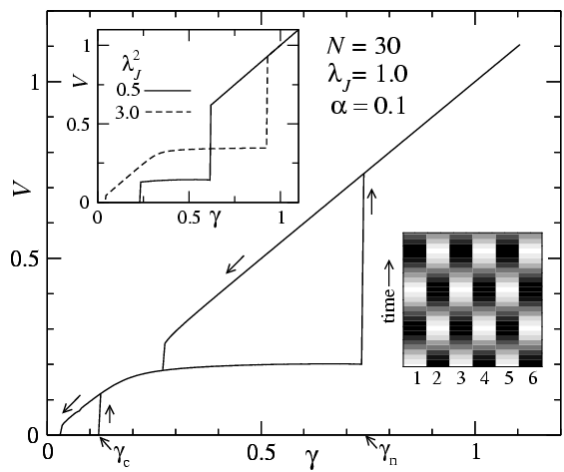} 
\caption{The full $V$-$I$ curve of the AFJJA. The direction of the current ramp is indicated 
by the arrow. Upper inset: the $V$-$I$ curves for different values of $\lambda_J$ (the 
$V$ branch with increasing $\gamma$ is shown). Left inset: the $2D$ space-time plot of 
$2\pi\frac{\Phi_{i}}{\Phi_{0}}$ in six cells of the array for $\gamma=0.3$, where the maximum 
value is 2.2 (black) and the minimum value is -2.2 (white).} 
\end{figure}

The origin of the plateau (or step) in $V$ can be understood from the fluxon dynamics. In a
0-JJA, the motion of a single $2\pi$-fluxon leads to the emission of small amplitude linear
waves (plasma waves) due to the discreteness of the array. The resonances between these linear
waves and the periodic motion of the fluxon causes a series of plateaus in
$V$~\cite{ustinov1}. Extending this to the case of AFJJA, the plateau in $V$-$I$ can be
attributed to the phase locking between the moving fluxon lattice and the linear waves emitted
by the array. The frequency $\omega_s$ of the linear waves can be calculated from Eq. (2). In
the absence of any external fluxon, the symmetry of the ground state allows only waves with
the wavevector $k\!=\!2\pi/(2a)$ to be coupled resonantly to the moving fluxon lattice (here
$a$ is the lattice constant). Thus, all 0 and $\pi$ junctions oscillate with the same
amplitude and phase. Linearizing Eq.(2) for the $\pi$ and the 0 junctions using
$u=u_{0}\exp(i\omega_{s}t)$ and $v=v_{0}\exp(i\omega_{s}t)$, respectively,
\begin{eqnarray}
-\omega_{s}^{2} u - u & = & 2\lambda_{J}^{2} (v - u), \nonumber \\
-\omega_{s}^{2} v + v & = & 2\lambda_{J}^{2} (u - v).
\end{eqnarray}
For simplicity, we have used $\alpha = \gamma = 0$. Adding the above equations gives the
relation between $u$ and $v$, $u = v (1-\omega_{s}^{2})/(1+\omega_{s}^{2})$. Substituting
for $u$ in the second equation leads to
\begin{equation}
(1-\omega_{s}^{2})(1+\omega_{s}^{2}) + 4\lambda_{J}^{2}\omega_{s}^{2} = 0.
\end{equation}
The above quadratic equation in $\omega_{s}^{2}$ can be solved to obtain the frequency of
the linear waves
\begin{equation}
\omega_{s}(\lambda_{J})=\sqrt{2\lambda_{J}^{2}+\sqrt{1+4\lambda_{J}^{4}}}.
\end{equation}
The $\omega_s$ is in units of the plasma frequency $\omega_P$. The condition for phase
locking of the linear waves with the moving fluxon lattice then becomes $\omega_{s} T =
2\pi$, where $T$ is the time period corresponding to the motion of the fluxon lattice. From
the simulation, the $T$ (and hence $\omega_{s}$) on the plateau can be obtained from the
time evolution of the magnetic flux $\Phi_{i}(t)$ in a cell. In Fig. 6(b), the $\omega_{s}$
calculated analytically is compared with that obtained from the simulation for increasing
$\lambda_J$. A reasonable agreement is observed over a range of $\lambda_J$. The
discrepancy appears as $\lambda_J$ approaches $\lambda_{J}^{*}$ below which the plateau in
$V$ is not observed. The voltage $V_s$ on the plateau is given by $V_{s}=\alpha\omega_{s}$.
This is shown in Fig. 6(c) and is in good agreement with the simulation.

\begin{figure}[hbt]
\includegraphics[width=210pt]{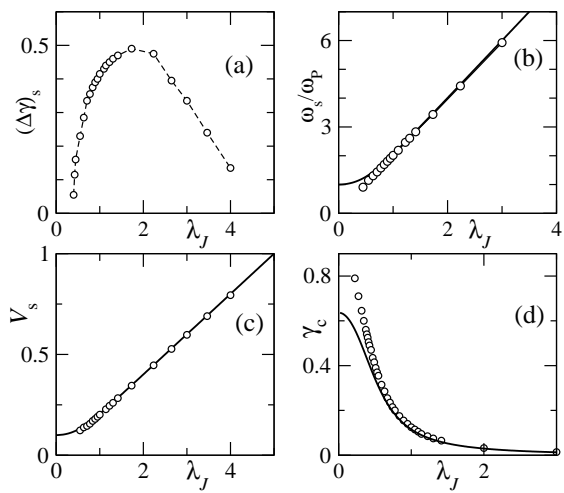} 
\caption{(a) The $\lambda_J$ dependence of the width of the voltage plateau
$(\Delta\gamma)_{s}$.  (b) Resonance frequency $\omega_{s}(\lambda_{J})$, and (c) the voltage
$V_{s}(\lambda_{J})$. (d) The depinning current $\gamma_c$. The open symbols are from the
simulation whereas the bold lines in (b), (c) and (d) are the analytical results described in
the text.} 
\end{figure} 

Finally, the dependence of $\gamma_{c}$ on the screening strength is shown in Fig. 6(d). A
finite $\gamma_c$ in JJA is attributed to the pinning potential created at the center of
the cell~\cite{lobb}. For AFJJA, the pinning potential or the energy barrier can be defined
as $\Delta E\!=\!E_{M}\!-\!E_{A}$, where $E_M$ is the energy of the array with the fluxon
lattice placed on the junctions and $E_A$ is the ground state energy. Equating the pinning
force $2\Delta E$ to the Lorentz force required to overcome the energy barrier gives the
depinning current $\gamma_c$ (per junction),
\begin{equation}
\gamma_c\!=\!\frac{1}{\pi}\left(2\cos(v)-\frac{1}{2}\lambda_{J}^{2}(\pi-2v)^{2}\right).
\end{equation} 
The phase $v$ in the above equation is the solution of the Eq.(5). Fig. 6(d) compares the
above expression with the $\gamma_{c}$ obtained from the simulation. The agreement is good
at large $\lambda_J$, and the deviation appears as $\lambda_{J}\rightarrow\lambda_{J}^{*}$
below which no flux flow is observed.

In conclusion, we have introduced a new class of JJA containing $\pi$ junctions. We considered
the one-dimensional case in which every alternate junction is a $\pi$ junction, and studied
its properties numerically and analytically. The ground state of the array contains
spontaneous magnetic flux in each cell and are ordered antiferromagnetically along the array.
A $2\pi$-fluxon in such an array is unstable and fractionalizes into two spatially separated
$\pi$-fluxons. The fractionalization is related to the ground state degeneracy. The $V$-$I$
curve shows a voltage plateau due to resonant interaction between the linear waves emitted by
the array and the moving antiferromagnetic fluxon lattice present in the ground state.

{\em Acknowledgments} : Part of this work was done during the authors stay at the University
of California, Davis.

{\em Note :} After the completion of the work, we became aware of the experimental
fabrication of an array of coupled $\pi$-loops using YBaCuO-Nb zigzag structure in
Ref.\cite{hilg}. Some of the results obtained in this paper are applicable for such
structures also. We thank the referees for bringing this paper to our attention.


\begin{references}

\bibitem{morpurgo} A. F. Morpurgo,T. M. Klapwijk, and B. J. van Wees, Appl. Phys. Lett. {\bf
72}, 966 (1998).

\bibitem{volkov} A. F. Volkov, Phys. Rev. Lett. {\bf 74}, 4730 (1995).

\bibitem{wilhelm} F. K. Wilhelm, G. Sch\"{o}n, and A. D. Zaikin, Phys. Rev. Lett. {\bf 81},
1682 (1998).

\bibitem{baselmans} J. J. A. Baselmans, A. F. Morpurgo, B. J. van Wees, and T. M. Klapwijk,
Nature {\bf 397}, 43 (1999); J. J. A. Baselmans, B. J. van Wees, and T. M. Klapwijk, Phys.
Rev. B {\bf 63}, 094504 (2001).

\bibitem{shaikh} R. Shaikhaidarov, A. F. Volkov, H. Takayanagi, V. T. Petrashov, and P.
Delsing, Phys. Rev. B {\bf 62}, R14649 (2000).

\bibitem{comment1} The $\pi$ junctions have been made also by coupling two superconductors
across a ferromagnetic layer, see V. V. Ryazanov, V. A. Oboznov, A. Yu. Rusanov, A. V.  
Veretennikov, A. A. Golubov, and J. Aarts, Phys. Rev. Lett. {\bf 86}, 2427 (2001).

\bibitem{baselmans1} J. J. A. Baselmans, B. J. van Wees, and T. M. Klapwijk,
cond-mat/0107292 (2001).

\bibitem{kusum} F. V. Kusmartsev, Phys. Rev. Lett. {\bf 69}, 2268 (1992). See also L. N.
Bulaevskii, V. V. Kusii, and A. A. Sobyanin, Pis'ma Zh. Eksp. Teor. Fiz. {\bf 25}, 314
(1977) [JETP Lett. {\bf 25}, 290 (1977)].

\bibitem{mints} R. G. Mints, Phys. Rev. B {\bf 57}, R3221 (1998); R. G. Mints and I. Papiashvili, 
Phys. Rev. B {\bf 62}, 15214 (2000).

\bibitem{gold} E. Goldobin, D. Koelle, and R. Kleiner, Phys. Rev. B {\bf 66}, 100508(R) (2002).

\bibitem{mahesh} Mahesh Chandran, (unpublished).

\bibitem{kulkarni} R. V. Kulkarni, E. Almaas, K. D. Fisher, and D. Stroud, Phys. Rev. B {\bf 62},
12119 (2000).

\bibitem{zant} H. S. J. van der Zant, F. C. Fritschy, W. J. Elion, L. J. Geerligs, and J. E.
Mooij, Phys. Rev. Lett. {\bf 69}, 2971 (1992). See also A. van Oudenaarden, S. J. K.
V\'{a}rdy, and J.  E.  Mooij, Phys. Rev. Lett. {\bf 77}, 4257 (1996) for localization, and
A. van Oudenaarden and J. E.  Mooij, Phys.  Rev. Lett. {\bf 76}, 4947 (1996) for superfluid
to Mott-insulator transition in Josephson junction array.

\bibitem{elion} W. J. Elion, J. J. Wachters, L. L. Sohn, and J. E. Mooij, Phys. Rev. Lett.
{\bf 71}, 2311 (1993).

\bibitem{barb} P. Barbara, A. B. Cawthorne, S. V. Shitov, and C. J. Lobb, Phys. Rev. Lett.
{\bf 82}, 1963 (1999).

\bibitem{barone} A. Barone and G. Paterno, {\it Physics and Applications of the Josephson
Effect}, (Wiley, New York, 1982).

\bibitem{ustinov1} A. V. Ustinov, M. Cirillo, and B. A. Malomed, Phys. Rev. B {\bf 47}, 8357
(1993).

\bibitem{ustinov2} A. V. Ustinov, B. A. Malomed, and S. Sakai, Phys. Rev. B {\bf 57}, 11691
(1998) and references there in.

\bibitem{comment3} We refer the fluxon created by applying a magnetic field to the array 
as an external fluxon. The external fluxon must be distinguished from the self-induced
fractional fluxons present in the ground state of the array which appears in the {\em 
absence} of an applied magnetic field.

\bibitem{comment2} For the simulation purpose, we consider $I_{c}^{0}=1$ and vary $I_{c}^{\pi}$.

\bibitem{su} W. P. Su, J. R. Schrieffer, and A. J. Heeger, Phys. Rev. Lett. {\bf 42}, 1698
(1979);  Phys. Rev. B {\bf 22}, 2099 (1980).

\bibitem{jackiw} R. Jackiw, and J. R. Schrieffer, Nucl. Phys. {\bf 190}[{\bf FS3}], 253 (1981).

\bibitem{rajaraman} R. Rajaraman, cond-mat/0103366, (2001).

\bibitem{comment4} The value of $\alpha$ chosen here may be small for the 3-terminal
$\pi$ junctions which have been implemented so far. We have verified that this does not
affect the qualitative features discussed in the paper. A more realistic case would be to
consider different values of $\alpha$ for 0 and $\pi$ junctions.

\bibitem{likharev} Breathing mode of a vortex-antivortex pair in long Josephson junction have 
been studied previously. See K. K. Likharev, {\it Dynamics of Josephson junctions and
circuits}, (Gordon and Breach, Amsterdam, 1986).

\bibitem{lobb} C. J. Lobb, D. W. Abraham, and M. Tinkham, Phys. Rev. B {\bf 27}, 150 (1983).
M. S.  Rzchowski, S. P. Benz, M. Tinkham and C. J. Lobb, Phys. Rev. B {\bf 42}, 2041 (1990).

\bibitem{hilg} H. Hilgenkamp, Ariando, Henk-Jan H. Smilde, D. H. A. Blank, G. Rijnders, H.
Rogalla, J. R. Kirtley, and C. C. Tsuei, Nature {\bf 422}, 50 (2003).

\end{references}
\end{document}